\documentclass[a4paper,12pt]{article}
\usepackage{epsfig}
\newcommand{\be}{\begin{equation}}
\newcommand{\ee}{\end{equation}}
\newcommand{\bear}{\begin{eqnarray}}
\newcommand{\ear}{\end{eqnarray}}
\newcommand{\pe}{\langle p^2_t\rangle}
\newcommand{\grgl}{\:\hbox to -0.2pt{\lower2.5pt\hbox{\small$\sim$}\hss}
                 {\raise3pt\hbox{$>$}}\:}
\newcommand{\klgl}{\:\hbox to -0.2pt{\lower2.5pt\hbox{\small$\sim$}\hss}
                 {\raise3pt\hbox{$<$}}\:}
\begin{document}

\begin{center}
{\Large \bf Time structure of anomalous {\boldmath $J/\psi$} and
            {\boldmath $\psi'$} suppression in nuclear collisions }\\[5mm]
      {J\"org H\"ufner$^{a}$, Pengfei Zhuang$^b$}\\[4mm]
      {\small{\it $^a$ Institute for Theoretical Physics,
                       University of Heidelberg, D-69120 Heidelberg,
                       Germany.} \\[1mm]
             {\it $^b$ Physics Department, Tsinghua University,
                       Beijing 100084, PR China}\\[5mm]}
\today\\[5mm]
\end{center}

\begin{abstract}
\setlength{\baselineskip}{16pt}
The data for the mean squared transverse momentum $\pe(E_t)$ as
function of transverse energy $E_t$ of $J/\psi$ and $\psi'$ produced
in Pb-Pb collisions
at the CERN-SPS are analyzed and it is claimed that they contain
information about the time structure of anomalous suppression.
A transport equation which describes transverse motion of $J/\psi$ and
$\psi'$ in the absorptive medium is proposed and solved for a
QGP and
a comover scenario of suppression. While the comover approach accounts
for the data fairly
well without adjusting any parameter, the fit to the data within the
QGP scenario requires to assume anomalous suppression to become
effective rather late, $3-4$ fm/c after the nuclear overlap.
\end{abstract}

\vspace{0.3in}

\setlength{\baselineskip}{18pt}

\begin{figure}[ht]
\vspace*{+0cm} \centerline{\epsfxsize=10cm  \epsffile{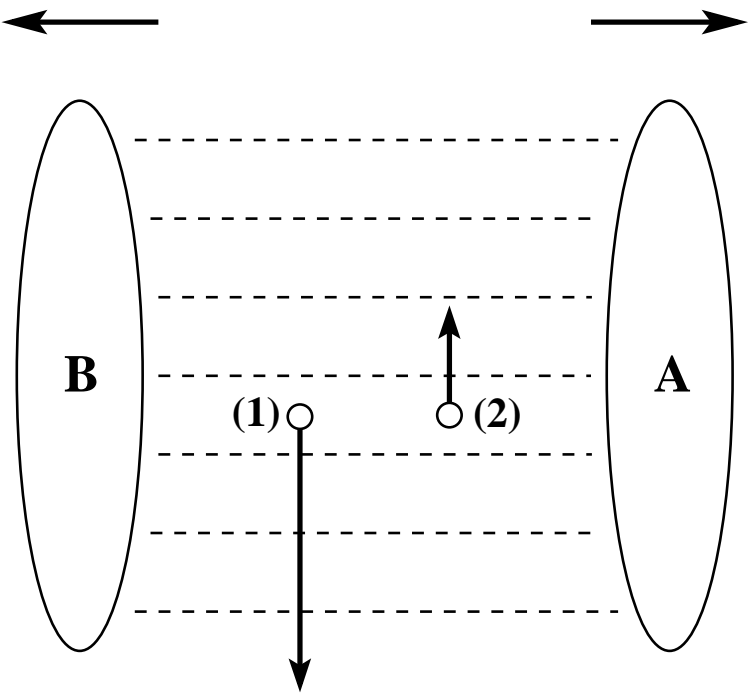}}
\caption{\it Schematic picture of the leakage phenomenon. Between
the
             Lorentz contracted remnants ``B'' and ``A'' of the nuclei
             which have collided charmonia
             move in the created parton/hadron plasma. Those $\psi$'s with
             large transverse velocities $v_t$ (case (1)) may leak
             out and escape suppression, while low $v_t$ particles
             may remain (case(2)) leading to an effective increase of
             $\langle p^2_t\rangle$ for the surviving (observed) $\psi$'s.}
\label{fig1}
\end{figure}

The discovery in 1996 of anomalous $J/\psi$ suppression in Pb-Pb collisions at
the SPS has been one of the highlights of the research with
ultrarelativistic heavy ions at CERN \cite{1}. Does it point to the
discovery of the predicted quark gluon plasma (QGP)? Six years later the
situation is still confused, since several models  - with and without
the assumption of a QGP - describe the observed suppression, after at
least one parameter is adjusted. The data on the mean squared
transverse momentum $\langle p^2_t\rangle(E_t)$ \cite{2} for the
$\psi$ (this symbol stands for $J/\psi$ and $\psi'$) in the regime of
anomalous suppression and as a function of transverse energy $E_t$
have received less attention - for no good reason. We claim: $\langle
p^2_t\rangle(E_t)$ contains additional information about the
nature of anomalous
suppression and may help to distinguish between different
scenarios. In this paper we investigate how the time structure of
anomalous suppression influences the values $\langle
p^2_t\rangle(E_t)$. This idea has
already been considered more than 10 years ago \cite{3} (c.f. also more
recent works~\cite{4,5}) and is based on the following observation
(Fig.\ref{fig1}): Anomalous suppression is not an instantaneous process, but
takes a certain time depending on the mechanism. During this time
$\psi$'s produced with high
transverse momenta may leak out of the parton/hadron plasma and escape
suppression. As a consequence, low $p_t$ $\psi$'s are absorbed preferentially
and the $\langle p_t^2 \rangle$ of the surviving
(observed) $\psi$'s show an increase $\delta \langle p_t^2\rangle$,
which grows monotonically with the mean time $t_A$, when anomalous
absorption acts \cite{5}. In this letter we propose
a general formalism of how to incorporate the effect of leakage into
the various
models, which have been proposed to describe anomalous suppression and
we extract information about the time $t_A$ from a comparison
with experiment.

It has become customary to distinguish between normal and anomalous
values of suppression $S(E_t)=\sigma^\psi(E_t)/\sigma^{DY}(E_t)$ for
$\psi$'s produced in
 nuclear collisions, c.f. reviews~\cite{6,7}. Here, $\sigma^\psi$ and
$\sigma^{DY}$ are the
production cross section for a $\psi$ and
a Drell-Yan pair in an AB collision, respectively.
By definition, $\psi$'s produced in $pA$ collisions show normal
suppression via inelastic $\psi N$ collisions in the final state  and
normal increase of $\pe$ (above $\pe_{NN}$ in $NN$ collisions) via
gluon rescattering in the initial state. These
normal effects are also present in nucleus-nucleus collisions and
happen, while projectile and target nuclei overlap. Anomalous values
of $S$ and $\pe$ are attributed to the action on the $\psi$
by the mostly baryon free phase of partons and/or hadrons (we call it
parton/hadron plasma) which is
formed  \underline{after} the nuclear overlap. It
may lead to deconfinement of the $\psi$ via colour screening in the QGP,
dissociation via gluon absorption
or inelastic collisions by the comoving hadrons during the later
period of
the plasma
evolution. In this letter we describe anomalous $\psi$ suppression
within a transport theory and apply it to two rather
different scenarios: (I) Absorption involving a threshold in the
energy density like in a QGP scenario, (II) continuous absorption via comovers.

We denote by $d\sigma^\psi/d\vec {p_t}(\vec p_t,E_t)$ the cross section
for the production of a $\psi$ with given $p_t$ and in an event with fixed
transverse energy $E_t$. It can be related to the phase space density
$f^\psi$ via
\be
\label{1}
\frac{d\sigma^\psi}{d\vec p_t}\,(\vec p_t,E_t)=
\lim_{t\rightarrow\infty} \int
d\vec b P(E_t;b)\int d\vec s \, f^\psi(\vec s,\vec p_t,t;\vec b).
\ee
Here, $P(E_t;b)$ describes the distribution of transverse energy $E_t$
in events with a given impact parameter $\vec b$ between projectile A
and target B. We follow ref. \cite{8}
in notation for $P(E_t;\vec b)$
and the values of the numerical constants. The function
$f^\psi(\vec s,\vec p_t,t;\vec b)$ is the distribution of $\psi$'s in the
transverse phase space $(\vec s,\vec p_t)$ at time $t$ for given $\vec b$.

We define $t=0$ as the time, when the process of normal suppression and normal
generation of $\pe$ has ceased and denote by $f^\psi_N(\vec s,\vec
p_t;\vec b)$ the distribution of $\psi$'s  at this time.
$f^\psi_N$ is taken as initial condition for the motion and absorption of the
$\psi$'s during the action of anomalous
interactions. The evolution of the $\psi$ is described
by a transport equation
\be
\label{2}
\frac{\partial}{\partial t} f^\psi + \vec
v_t\cdot\vec\nabla _s f^\psi=-\alpha f^\psi.
\ee
The time dependence arises from the free streaming of the $\psi$ with
transverse velocity $\vec v_t=\vec p_t/\sqrt{m^2_\psi+p^2_t}$ (l.h.s.)
and an absorptive term on the r.h.s., where the function
$\alpha(\vec s,\vec p_t,t;\vec b)$
contains all details about the surrounding matter and the absorption
process. We have left out effects from a mean field, because the
elastic $\psi N$ cross section is very small and have also neglected a
gain term  on the r.h.s., because recombination processes $c+\bar
c+N\to\psi+X$ seem unimportant at SPS energies where at most one
$c\bar c$ is created per event.

Eq.~(\ref{2}) can be solved analytically with the result
\be
\label{3}
f^\psi(\vec s,\vec p_t,t;\vec b)=\exp\left(-\int^t_0 dt'\alpha(\vec
s-\vec v_t(t-t'),\vec p_t,t';\vec b)\right)f^\psi_N(\vec s-\vec v_t
t,\vec p_t;\vec b),
\ee
which for $t=0$ reduces to $f^\psi=f^\psi_N$. If we denote by $t_f$
the time when anomalous suppression has ceased, $\alpha(\vec
s, \vec p_t,t;\vec b)=0$ for $t>t_f$, the limit $t\rightarrow\infty$
in eq.~(\ref{1}) can be replaced by setting $t=t_f$, since the distribution in $p_t$
does not change for larger $t$'s.

There is little controversy about $\psi$ production and suppression
in the normal
phase: The gluons, which fuse to the $c\bar c$, collide with nucleons
before fusion and gain additional $p_t^2$. The $\psi$ on its way out
is suppressed by inelastic $\psi N$ collisions without any change in
$\langle p^2_t\rangle$. Neglecting effects of
formation time\cite{9}, one has
\bear
\label{4}
f^\psi_N(\vec s,\vec p_t;\vec b)
&=&
\sigma_{NN}^\psi \int dz_A\, dz_B\, \rho_A(\vec s,z_A)
\rho_B(\vec b-\vec s,z_B)\cdot\nonumber\\
& & \cdot\exp\left(-\sigma_{abs}^\psi
[T_A(\vec s,z_A,\infty)+T_B(\vec b-\vec s,-\infty,z_B)]\right)
\pe_N^{-1}\exp\left(-p_t^2/\pe_N\right),
\ear
where
\be
\label{5}
\pe_N(\vec b,\vec s, z_A,z_B)=\pe^\psi_{NN}+a_{gN}\rho^{-1}_0[T_A(\vec
s,-\infty,z_A) +T_B(\vec b-\vec s,z_B,+\infty)]
\ee
with the thickness function $T(\vec s,z_1,z_2)=\int^{z_2}_{z_1}dz
\,\rho(\vec s,z)$.
All densities $\rho_A,\rho_B$ are normalized to the number of nucleons
($\rho_0$ is the nuclear matter density). We shortly explain
eqs.~(\ref{4}) and (\ref{5}): For given values $\vec b$ and $\vec s$ in the
transverse plane the $\psi$ is produced at coordinates
$z_A$ and $z_B$ in nuclei
$A$ and $B$, respectively. On its way out, the $\psi$ experiences the
thickness $T_A(\vec s,z_A,\infty)$ and $T_B(\vec b-\vec
s,-\infty,z_B)$ in nuclei $A$ and $B$, respectively and is suppressed
with an effective absorption cross section $\sigma^\psi_{abs}$. The
two gluons which fuse carry transverse momentum from two sources: (i)
Intrinsic $p_t$, because they had been confined to a nucleon. The
intrinsic part is observable in $NN\to\psi$ collisions and leads to
$\pe^\psi_{NN}$ in eq.~(\ref{5}).
(ii) Collisional contribution to $p_t$, because in a nuclear
collision, the gluons traverse thicknesses $T_A(\vec s,-\infty,z_A)$
and $T_B(\vec b-\vec s,z_B,+\infty)$ of nuclear matter in A and B,
respectively,
and acquire additional transverse momentum via $gN$ collisions. This is the
origin of the second term in eq.~(\ref{5}).

The constants $\sigma^{J/\psi}_{abs},\sigma^{\psi'}_{abs}$ and
$a_{gN}$ are usually adjusted to the data from $pA$ collisions, before
one investigates anomalous suppression. Fig.~2 shows (dashed
curves) the results for normal suppression $S^\psi(E_t)$ and
$\pe^\psi(E_t)$ calculated with $f_N^\psi$
eq.~(\ref{4}) in eq.~(\ref{1}). While the
difference between calculation and data is enormous for the
suppression, it is rather small for $\pe^\psi(E_t)$.

Since the physical origin of anomalous suppression is not yet
settled, we investigate suppression $S^\psi(E_t)$
and $\pe^\psi (E_t)$ for two models, which
have rather contradictory assumptions.
\begin{enumerate}
\item[I.] Threshold (QGP-) scenario: $\psi$'s are totally and rapidly
destroyed, when
they are in a medium with energy density above a critical one, and
nothing happens elsewhere. As a representative model we use the
approach by Blaizot et al.~\cite{8}.
\item[II.]  Comover scenario: The plasma of comoving partons and/or
hadrons leads to a continuous absorption of long duration due to
inelastic collisions with the comoving
particles. As a representative model, we use the approaches by Capella et al.
\cite{10} and Kharzeev  et al. \cite{11}.
\end{enumerate}

In this letter we study the effect of leakage on the observed values
of $\langle p^2_t\rangle$ within two well established scenarios, using
their assumptions and parameters. We do not introduce any
modifications like (i) the $p_t$ dependence of the absorption process
(i.e. $p_t$ dependence of $\alpha(\vec s,\vec p_t,t;b)$ in
eq.~(\ref{2})) and (ii) expansion of the plasma during
absorption. Both effects may contribute to $\langle p^2_t\rangle$ (we
will present a qualitative discussion at the end), but both require a
detailed study by themselves.

We begin with model I: In their schematic approach
Blaizot et al.~\cite{8} include anomalous suppression via
\be
\label{6}
f^\psi(\vec s,\vec p_t;\vec b)=\theta(n_c-n_p(\vec b,\vec
 s))f^\psi_N(\vec s,
\vec p_t;\vec b).
\ee
Here $n_c$ is a critical density and $n_p(\vec b,\vec s)$ is
the density of participant nucleons
\be
\label{7}
n_p(\vec b,\vec s)=T_A(\vec
s,-\infty,+\infty)[1-\exp\left(-\sigma^{NN}_{in}T_B(\vec b-\vec
s,-\infty,+\infty)\right)]+(A\leftrightarrow B).
\ee
According to eq.~(\ref{6}) all $\psi$'s are destroyed if the energy
density (which is directly proportional to the participant density) at
the location $\vec s$ of the $\psi$ is larger than the critical
density. All other $\psi$'s survive.
While
the prescription eq.(\ref{6}) successfully describes the data in the full
$E_t$ range of anomalous suppression after the only one free parameter,
$n_c$, is adjusted, the predictions for $\pe(E_t)$
are significantly below the data, especially at large $E_t$ (see below).

The expression eq.~(\ref{6}) for the phase space distribution $f^\psi$
including anomalous suppression within the threshold model is
recovered within our transport approach eq.~(\ref{3}) by setting
\be\label{8}
\alpha(\vec s,\vec p,t;\vec b)=\alpha_0\theta(n_p(\vec b,\vec
s)-n_c)\delta(t)
\ee
and taking the limit $\alpha_0\to\infty$. The delta function
$\delta(t)$ has to be included to recover the expression eq.~(\ref{6})
and may be understood by the physical picture
that the  energy density is highest at $t=0$
and anomalous absorption most effective then.

There are various ways to introduce another time structure into
the absorption term. We will try two options, one being \be
\label{9} \alpha(\vec s,\vec p_t,t;\vec b)=\alpha_0\theta
(n_p(\vec b,\vec s)-n_c)\delta(t-t_A). \ee The idea of a threshold
density is kept, but suppression does not act at $t=0$ but at a
later time $t_A$, which time is then determined from a comparison
with the data. For  times $t>t_A$, one finds from the general
solution eq.~(\ref{3}) (and $\alpha_0\to\infty$)
\be \label{10}
f^\psi(\vec s,\vec p_t,t)=\theta(n_c-n_p(\vec b,\vec
s)) f^\psi_N(\vec s-\vec v_t t_A,\vec p_t),\quad t>t_A
\ee
which differs from the expression (\ref{6}), by the motion in phase
space of $f^\psi_N$ during the time $ 0\leq t\leq t_A$. For $t>t_A$
the momentum distribution derived from $f^\psi$ does
not change any more.

The suppression $S^\psi(E_t)$ and $\pe\psi (E_t)$ calculated with
the distribution  $f^\psi$ from eq.~(\ref{10}) depends on the
parameter $t_A$. We use the parameters of ref.~\cite{8} where
available, i.e. for $J/\psi$:  $\sigma^{J/\psi}_{abs}=6.4$ mb,
$n_c = 3.75$ fm$^{-2}$. The parameters for the generation of $\pe$
by gluon rescattering are taken from a fit to the $pA$
data~\cite{2}: $\pe_{NN}=1.11$ (GeV$/c)^2$ and $a_{gN}=0.081$
(GeV$/c^2)$ fm$^{-1}$. We also account for the transverse energy
fluctuations~\cite{8} which have been shown to be significant for
the explanation of the sharp drop of $J/\psi$ suppression in the
domain of very large $E_t$ values, by replacing $n_p$ by
${E_t\over \langle E_t\rangle} n_p$ where $\langle E_t\rangle$ is
the mean transverse energy at given $b$. We then calculate
$\sigma^{J/\psi}/\sigma^{DY}$ as a function of $t_A$. Since the
critical density for $J/\psi$ is quite high, the leakage affects
only the very high momentum $\psi$'s. Since their number is small,
the calculated results for the suppression $S^{J/\psi}(E_t)$
depend only little on $t_A$ (Fig. 2).

We turn to a discussion of $\pe^{J/\psi}(E_t)$. Fig.~2 shows
calculated curves for values of $t_A=0$ to 4~fm/$c$. The dotted line
($t_A=0$) is the result of the original threshold
model with immediate anomalous suppression (and has been
predicted in \cite{12}). It fails badly at large values of
$E_t$. Also no other curve with a given $t_A$ describes the data for
all values of $E_t$. We have to conclude that $t_A$ depends on $E_t$:
$t_A(E_t)$. The larger the values $E_t$ the later anomalous
suppression acts. From a comparison with data we have
$t_A\klgl2$~fm$/c$ for $E_t<80$~GeV, $t_A\simeq2.5$~fm$/c$ for $80\leq
E_t\leq 100$~GeV, and $t_A\simeq 3.5$~fm/c for $E_t>100$~GeV.

\begin{figure}[ht]
\hspace{+0cm} \centerline{\epsfxsize=10cm  \epsffile{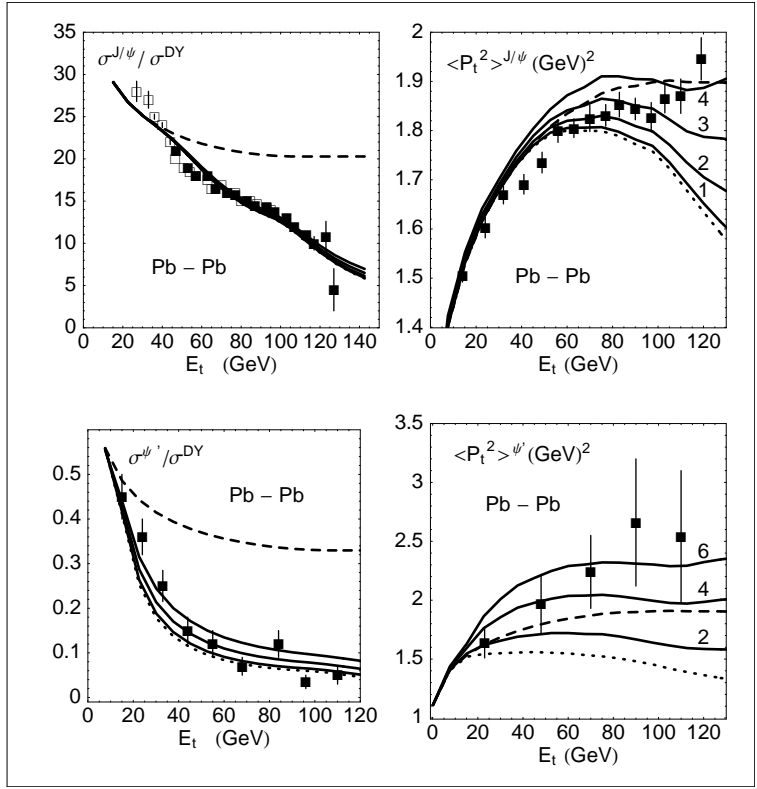}}
\caption{\it Nuclear suppression $\sigma^\psi/\sigma^{DY}$ and
$\pe^\psi$ for $J/\psi$ (above) and $\psi'$ (below) as a function
of transverse energy $E_t$.  Data are from \cite{1} for
$\sigma^\psi/\sigma^{DY}$ and from \cite{2} for $\pe$. Dashed
curves show the result of normal suppression alone. The dotted
lines correspond to the original threshold model, which in our
notation corresponds to $t_A=0$. The other curves include
anomalous suppression within the threshold model eqs.(\ref{6}) and
(\ref{10}), where anomalous suppression is assumed to act at time
$t_A>0$. The curves are labeled by the values $t_A =1,2,3,4 $fm/c
for $J/\psi$ and $t_A = 2,4,6$ fm/c for $\psi'$. Also the curves
in $\sigma^\psi/\sigma^{DY}$ carry these labels, lowest curve $t_A
= 0$ and monotonic increase with $t_A$. } \label{fig2}
\end{figure}

\begin{figure}[ht]
\hspace{+0cm} \centerline{\epsfxsize=10cm  \epsffile{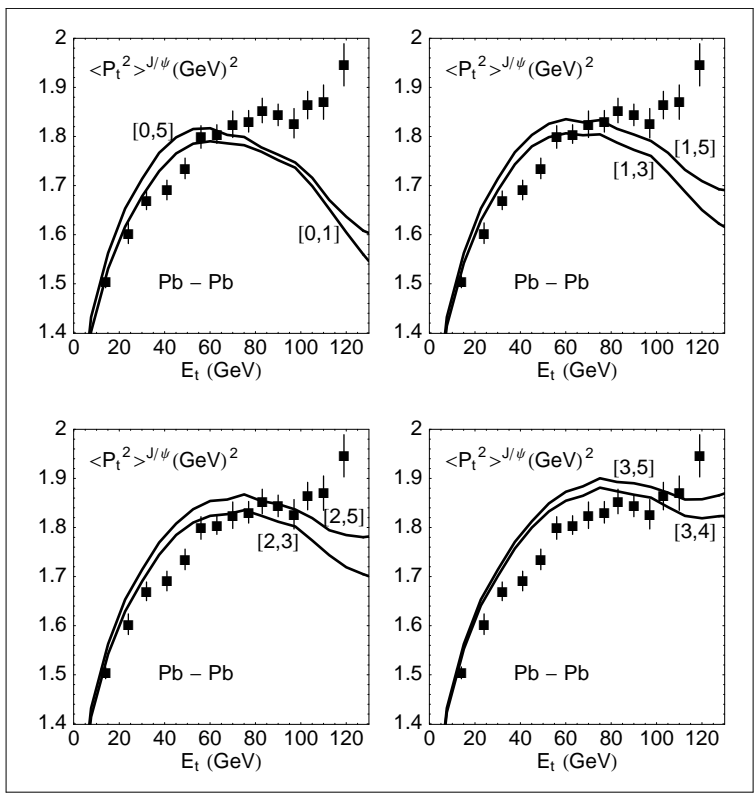}}
\caption{\it $\pe$ for $J/\psi$ as a function of transverse energy
$E_t$.  Data are from \cite{2} for $\pe$. The solid lines are
calculated within the threshold model but with a continuous
suppression in the time interval $[t_0,t_1]$, c.f.
eq.~(\ref{11}).} \label{fig3}
\end{figure}

A similar analysis is performed for the $\psi'$. The data for
the suppression are  taken from~\cite{13,14}, those for $\pe^{\psi'}$
 from~\cite{2}. Since the $\psi'$ has not been treated by Blaizot et
al., we fit the suppression data and  find values for the absorption
parameters,
$\sigma^{\psi'}_{abs}=7$ mb, $n_c^{\psi'}=2.3$ fm$^{-2}$, leaving the
parameters $\pe_{NN}$ and
$a_{gN}$ unchanged. Since the critical density for $\psi'$ is
smaller than that for $J/\psi$, more $\psi'$s leak out of the
anomalous suppression region, the change in suppression due to the
increase of $t_A$ is noticeable. Fig. 2 shows the results
with a good fit to the
suppression data.

The data for $\pe^{\psi'}$ have rather large error bars. The
calculated curves show a strong dependence on $t_A$, again with a
trend that $\pe$ at larger values of $E_t$ require larger values of
$t_A$. However, the numerical values $t_A(E_t)$ for $\psi'$ are above
those for the $J/\psi$ by about $1-2 fm/c$. This result is
strange, because we expect the $\psi'$ to be destroyed more easily
and therefore more rapidly.  We will come back to this point
in the conclusions.

The time structure introduced via eq.~(\ref{9}) is certainly
oversimplified. Rather than having it act at one particular time
$t_A$, it is more reasonable to assume that it acts during a time
interval. Therefore we have also investigated the following form of
the absorptive term
\be
\label{11}
\alpha(\vec s,\vec p_t,t;\vec b)={\alpha_0\over  t_1-t_0}\theta
(n_p(\vec b,\vec
s)-n_c)\theta(t_1-t)\theta(t-t_0).
\ee
Anomalous suppression acts for times $t$ between $t_0$ and $t_1$
(which may be a function of $E_t$).
Especially $t_0$ is an interesting quantity, because it
gives the starting time for anomalous suppression.

We calculate the suppression $S^{J/\psi}(E_t)$ and $\pe^{J/\psi}(E_t)$
as a function of $[t_0,t_1]$ and for $\alpha_0=5$. All other
parameters are unchanged. The suppression is well described for all intervals
$[t_0,t_1]$, but the calculations for $\pe^{J/\psi}(E_t)$ depend
strongly on the choice of this interval. Fig.3 shows some
representative examples: The four windows in Fig.3 display events
with $t_0 =0,1,2,3 fm/c$, respectively, and several values for
$t_1$. It is obvious that there is not one curve, which describes all
the data. Rather we find that the best curve for $E_t < 60$ GeV corresponds
to $[0,1]$, for $60 < E_t < 80$ GeV to $[1,3]$, for $80 < E_t < 100$ GeV
to $[2,5]$, and for $E_t > 100$ GeV to $[3,5]$. The length $t_1 - t_0$ of
the interval remains approximately constant, while the beginning time $t_0$
increases with  $E_t$.  We will discuss the significance of these
 results after we have treated leakage within the comover model.

We proceed to model II, the scenario of comovers: Partons and/or
hadrons which move with the $\psi$ may destroy the $\psi$ with a
cross section $\sigma^\psi_{co}$. The comover density $n_{co}(t)$
depends on time, for which the Bjorken  scenario of longitudinal
expansion predicts $t^{-1}$. The absorptive term $\alpha$ in
eq.~(\ref{2}) then takes the form \be \label{12} \alpha(\vec
s,\vec p_t,t;\vec b)=\sigma^\psi_{co}{n_c(\vec b,\vec s)\over
t}\theta({n_c(\vec b,\vec s)\over n_f}t_0 - t)\theta(t-t_0), \ee
i.e. absorption by comovers starts at $t=t_0$ and ends at $t_1$,
when the comover density $n_c(\vec b,\vec s)\cdot t_0/t_1$ has
reached a value $n_f$ independent of $\vec b$ and $\vec s$. The
comover approach contains a definite time structure for anomalous
suppression and we have not changed it. We also account for the
transverse energy fluctuations\cite{10}, by replacing $n_c$ by
${E_t\over <E_t>}n_c$, and the transverse energy loss\cite{10a}
induced by the $J/\psi$ trigger, by rescaling $<E_t>$ by a factor
${n_p -2 \over n_p}$, which have been shown to be significant for
the explanation of the sharp decrease of $S^{J/\psi}(E_t)$ at $E_t
> 100$ GeV. In the choice of parameters we have followed \cite{10,11}:
$n_{co}(\vec b,\vec s)=1.5\ n_p(\vec b,\vec s)$ with the
participant density from eq.~(\ref{7}), $t_0=1$~fm/c,\
$n_f=1$~fm$^{-2}$,\ $\sigma^{J/\psi}_{abs}=4.5$~mb,
$\sigma^{\psi'}_{abs}=6$~mb, $\sigma^{J/\psi}_{co}=1$~mb and
$\sigma^{\psi'}_{co}=3$~mb, no additional parameter has been
introduced.

The calculated $E_t$ dependence of the suppression shown in Fig.4
fits the data acceptably well like model I. For $\pe$, we compare
the results of two calculations with the data. The dashed lines in
Fig.~4 show the calculation leaving out leakage. Formally this
limit  is  obtained from eq.~(\ref{3}) by setting $\vec v_t =0$ in
the exponent and in $f_N^\psi$. Due to the introduction of the
$E_t$ loss which is necessary to recover the $J/\psi$ suppression
for large values of $E_t$, the case without leakage does not fit
the data $\pe$ even in the domain of low $E_t$ values. Only when
the leakage effect is taken into account, the calculation agrees
well with the data. We stress: the calculation of $\pe (E_t)$ in
the comover model is a true prediction in the sense that no
parameter is adjusted above those which are fitted to the
suppression $S^\psi(E_t)$. We have also calculated the mean time
$\langle t_A^\psi\rangle$ for comover action by studying the
suppression $S^\psi(E_t)$ as a function of time and taking the
mean of $t$ with the weight $d S^\psi(E_t)/dt$ and find \bear
\label{13}
\langle t^{J/\psi}_A\rangle &=& 3.5\ {\rm fm}/c  \nonumber \\
\langle t^{\psi'}_A\rangle  &=& 3.0\ {\rm fm}/c,
\ear
which values include the time $t_0$, eq.~(\ref{12}), between the end
of normal suppression and the beginning of comover action. The values
eq.~(\ref{13}) are found to be rather independent of $E_t$.

\begin{figure}[ht]
\hspace{+0cm} \centerline{\epsfxsize=10cm  \epsffile{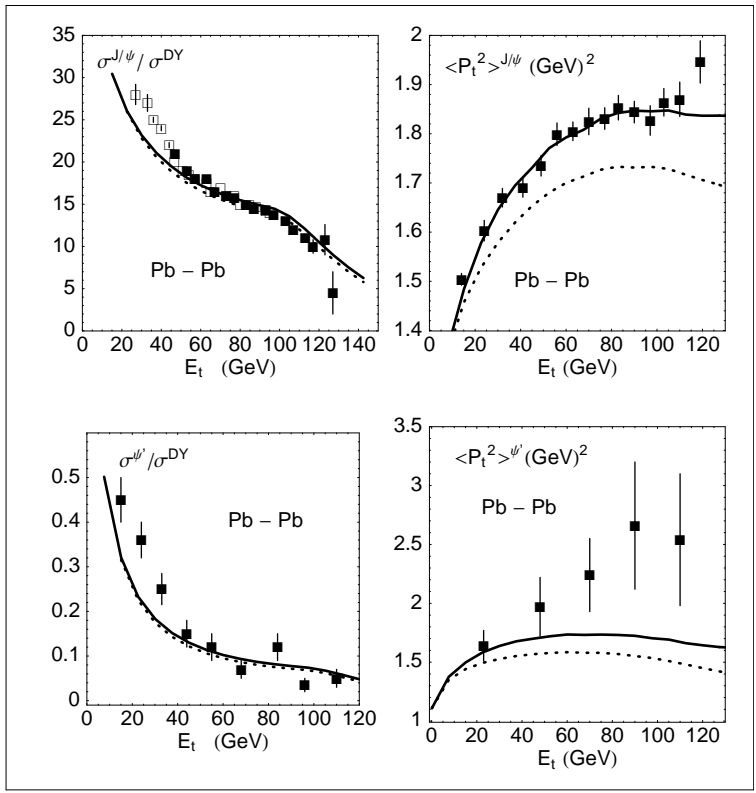}}
\caption{\it Nuclear suppression $\sigma^\psi/\sigma^{DY}$ and
$\pe$ for $J/\psi$ (above) and $\psi'$ (below) as a function of
transverse energy $E_t$. Dotted and solid lines are calculated in
the comover model without and with considering the leakage effect,
respectively. } \label{fig4}
\end{figure}

We summarize: In this letter we have investigated the influence of
leakage on the  calculation of
$\langle p^2_t\rangle(E_t)$ for $J/\psi$ and $\psi'$
produced in Pb-Pb collisions at SPS energies. This effect is closely
related to  time structure of anomalous suppression.
 The  evolution of
the $\psi$ during anomalous suppression including leakage is described
within a general transport equation. The formalism is applied to two
models with rather contradictory underling physical assumptions, the threshold
 (QGP) and the comover models with the following results:
\begin{enumerate}
\item[(i)] Calculations within the original models, where leakage is left out,
do not describe the data for $\pe^\psi(E_t)$, the discrepancy
being particularly strong at high values of $E_t$.

\item[(ii)] Including leakage into the comover model, without changing
  its
structure
nor its parameters leads to a good agreement with the data for $\pe(E_t)$
for $J/\psi$ over the full range of values $E_t$.

\item[(iii)] The assumption in the threshold model that anomalous suppression
acts instantaneously at $t_A = 0$, i.e. right after normal suppression is
not supported by experiment. Rather for central collisions, the data
of $\pe$ are described best, if one assumes that anomalous suppression
acts at a time $t_A=3-4$ fm$/c$ after the nuclear overlap.

\item[(iv)] The situation for the $\psi'$ is less clear. While the data for
suppression can be fitted by properly adjusting the parameters, both models
underpredict the data for $\pe^{\psi'}(E_t)$. This is evident for the
comover model. Within the threshold model the values $t_A$ required to fit
the $\psi'$ data are larger than those for the $J/\psi$, which seems
unreasonable to us. It could be that the error bars on the experiments
are too small.
\end{enumerate}

We conclude: In this letter we have investigated how leakage
(escape of high $p_t\, \psi'$s) when introduced into exciting
models of anomalous suppression influences the calculated values
of $\pe (E_t)$. As mentioned already above, there could also be
other effects which could influence $\pe$. We discuss them
briefly: $p_t$ dependence of the mechanism responsible for
anomalous suppression and transverse expansion of the plasma. Both
effects can be treated within the transport approach
eqs.~(\ref{2}),(\ref{3}) by introducing an explicit dependence on
$p_t$ and a modified dependence on $t$ into the function
$\alpha(s,p_t,t;b)$.   A first and schematic investigation  on the
$p_t$  has been made by  Dingh~\cite{17}.  One sees without
calculation that if  $\partial\alpha/\partial p_t<0$, i.e. high
$p_t\,\psi'$s are absorbed  less (whatever the mechanism may be),
the  calculated values of $\pe (E_t)$ increase, thus having the
same effect as an increase of the time $t_A$ eq.~(\ref{9}), when
anomalous suppression acts. As for the transverse expansion of the
plasma, its effect is qualitatively clear: It reduces the effect
of leakage since it makes it harder for the $\psi'$s to escape.
Although the qualitative changes on $\pe$ of the two effects are
clear, their  quantitative treatment needs a rather careful study
of the underlying physics, necessitates to modify existing models,
to change their input parameters and to introduce new parameters.
This is beyond the scope of this letter.

{\bf Acknowledgments}: We are grateful to our friends and colleagues
A. Gal, J. Dolej\v s\i, Yu. Ivanov, B.Z. Kopeliovich, H.J. Pirner and
C.Volpe for help and valuable
comments. One of the authors (P.Z.) thanks for the hospitality at the
Institute of Theoretical Physics. This work has been supported by the
grant 06HD954 of the German Federal Ministry of Science and Research
and by the Chinese National Science Foundation.


\begin{thebibliography}{20}

\bibitem{1} M.C. Abreu et al., Nucl. Phys. {\bf A 610} (1996) 404c.
\bibitem{2} M.C. Abreu et al., NA50 Collaboration, Phys. Lett. {\bf B
            499} (2001) 85.
\bibitem{3}F. Karsch, R. Petronzio, Z. Phys. {\bf C 37} (1988) 627;\\
           T. Matsui, H. Satz, Phys. Lett. {\bf B 178} (1986) 416;\\
           J.P. Blaizot, J.Y. Ollitrault, Phys. Lett. {\bf B 199} (1987) 627.
\bibitem{4} D. Kharzeev, M. Nardi, H. Satz, Phys. Lett. {\bf B 405}
           (1997) 14;\\
            E. Shuryak, D. Jeaney, Phys. Lett. {\bf B 430} (1998) 37.
\bibitem{5} J. H\"ufner, P. Zhuang, Phys. Lett. {\bf B 515} (2001)
            115.
\bibitem{6} R. Vogt, Phys. Rep. {\bf 310} (1999) 197.
\bibitem{7} C. Gerschel, J. H\"ufner, Annu. Rev. Nucl. Part. Sci. {\bf
            49} (1999) 255.
\bibitem{8} J.P. Blaizot, P.M. Dinh, J.Y. Ollitrault,
            Phys. Rev. Lett. {\bf 85} (2000) 4010;\\
            J.P. Blaizot, J.Y. Ollitrault, Phys. Rev. Lett. {\bf 77}
            (1996) 1703.
\bibitem{9} J. H\"ufner, B.Z.Kopeliovich, Phys. Rev. Lett. {\bf 76}
             (1996) 192.
\bibitem{10} A. Capella, E.G. Feirreiro and A.B. Kaidalov,
            Phys. Rev. Lett. {\bf 85} (2000) 2080.
\bibitem{10a} A. Capella, A. B. Kaidalov and D. Sousa,
            Phys. Rev. {\bf C65} (2002) 054908.
\bibitem{11} D. Kharzeev, C. Lovrenco, M. Nardi, H. Satz,
            Z. Phys. {\bf C} (1997).
\bibitem{12} D. Kharzeev, M. Nardi, H. Satz, Phys. Lett. {\bf B 405}
            (1997) 14.
\bibitem{13} M.C. Abreu et al., Phys. Lett. {\bf B 477} (2000) 28.
\bibitem{14} M.C. Abreu et al., Nucl. Phys. {\bf A 638} (1998) 261c.
\bibitem{15} N. Armesto, A. Capella, E.G. Feirreiro, Phys. Rev. {\bf
             C 59} (1999) 395.
\bibitem{16} D. Rischke, M. Gyulassy, Nucl. Phys. {\bf A 597} (1996) 701.
\bibitem{17} Private communication by J.Y. Ollitrault.

\end{thebibliography}
\end{document}